# A distance to the Large Magellanic Cloud that is precise to one per cent

G. Pietrzyński[1,2], D. Graczyk[1,2,3], A. Gallenne[4,5], W. Gieren[2], I. B. Thompson[6], B. Pilecki[1], P. Karczmarek[7], M. Górski[2], K. Suchomska[7], M. Taormina[1], B. Zgirski[1], P. Wielgórski[1], Z. Kołaczkowski[1,8], P. Konorski[7], S. Villanova[2], N. Nardetto[5], P. Kervella[9], F. Bresolin[10], R. P. Kudritzki[10,11], J. Storm[12], R. Smolec[1] & W. Narloch[1]

1. Nicolaus Copernicus Astronomical Centre, Warsaw, Poland.
2. Universidad de Concepción, Departamento de Astronomìa, Concepciòn, Chile.
3. Millennium Institute of Astrophysics (MAS), Santiago, Chileⓔ
4. European Southern Observatory, Santiago, Chile.
5. Université Côte d'Azur, Observatoire de la Côte d'Azur, CNRS, Laboratoire Lagrange, Nice, France.
6. Carnegie Observatories, Pasadena, CA, USA.
7. Warsaw University Observatory, Warsaw, Poland.
8. Astronomical Institute, Wrocław University, Wrocław, Poland.
9. LESIA, Observatoire de Paris, Université PSL, CNRS, Sorbonne Université, Univ. Paris Diderot, Sorbonne Paris Cité, Meudon, France.
10. Institute for Astronomy, Honolulu, HI, USA.
11. Munich University Observatory, Munich, Germany.
12. Leibniz Institute for Astrophysics, Potsdam, Germany.

**In the era of precision cosmology, it is essential to empirically determine the Hubble constant with an accuracy of one per cent or better[1]. At present, the uncertainty on this constant is dominated by the uncertainty in the calibration of the Cepheid period – luminosity relationship[2,3] (also known as Leavitt Law). The Large Magellanic Cloud has traditionally served as the best galaxy with which to calibrate Cepheid period-luminosity relations, and as a result has become the best anchor point for the cosmic distance scale[4,5]. Eclipsing binary systems composed of late–type stars offer the most precise and accurate way to measure the distance to the Large Magellanic Cloud. Currently the limit of the precision attainable with this technique is about two per cent, and is set by the precision of the existing calibrations of the surface brightness – colour relation[5,6]. Here we report the calibration of the surface brightness-colour relation with a precision of 0.8 per cent. We use this calibration to determine the geometrical distance to the Large Magellanic Cloud that is precise to 1 per cent based on 20 eclipsing binary systems. The final distane is 49.59 ± 0.09 (statistical) ± 0.54 (systematic) kiloparsecs.**

In order to provide an improved calibration of the relation between surface brightness and colour, we have carefully selected a sample of 41 nearby red clump giant stars, which are in the core helium burning phase of stellar evolution. We note that surface brightness $S_V$ is defined as:

$$S_v = V_0 + 5\log(\varphi) \quad (1)$$

where $V_0$ is the V band magnitude corrected for the reddening and $\varphi$ is the stellar angular diameter. We made sure that our sample does not contain variable stars or binaries. For our sample stars we collected precise near-infrared photometry at the South African Astronomical Observatory[7], and angular diameters precise to 1% with the ESO VLTI and PIONIER instrument[8]. These data are complemented with high-quality homogeneous V band photometry[9]. More details about the data are given in Methods.

The following relation was obtained from a least-squares fit:

$$S_V = (1.330 \pm 0.017) \times [(V-K)_0 - 2.405] + (5.869 \pm 0.003) \text{ mag} \quad (2)$$

with a root-mean-square (r.m.s.) scatter of 0.018 mag. Here V and K are V band and K band magnitude, respectively, and $(V - K)_0$ is the colour corrected for the extinction. Details about the extinction correction and the fit are given in Methods. Figure 1 shows the data together with the fit to a straight line, and the residuals from the relation. A comparison with the relation of Di Benedetto 6 is shown in Extended Data Fig. 2. Our new relationship allows a measurement of stellar angular diameters with a precision of 0.8% (that is, by a factor of 2.5 more precise than existing relations in the literature) 6 . It is perfectly suited to determine very accurate distances to eclipsing binaries composed of helium-burning giant stars.

We applied our new relation to determine the distances to 20 long-period, fully detached eclipsing binary systems in the Large Magellanic Cloud (LMC). All these extremely rare binaries are composed of helium-burning giants and were very carefully selected from the catalogue of 35 million stars observed over 20 years by the OGLE Project[10,11]. In the course of the Araucaria Project[12], we have collected high-quality spectroscopic data with the MIKE, HARPS and UVES high-resolution spectrographs, and near-infrared photometry with the SOFI camera at ESO-La Silla, and used these data to determine very precise astrophysical parameters for these systems (1%–3% masses, radii, temperatures, and so on[5,13]). The distances $d$ (in pc) were calculated from the equation $d = 9.2984 \times R/\varphi$, where the angular diameters $\varphi$ (in mas) given by $\varphi = 10^{0.2(S_V - V_0)}$ were calculated from equation (2), and V and K band magnitudes, and the radii R (in solar radii) were taken from ref. [13]. Realistic statistical errors on the distances were obtained from extensive Monte Carlo simulations taking into account all potential contributions (see Extended Data Fig. 3 and Extended Data Table 4 for more information). The distance results are presented in Table 1.

The weighted mean distance modulus of the LMC from our sample of 20 binary systems is 18.476 ± 0.002 mag. Since all targets are located relatively close to the centre of the LMC and to its line of nodes (see Fig. 2), this determination should only very weakly depend on the geometrical extension of this galaxy. In order to check this, we adopted a geometrical model of the LMC 14 and calculated the corrections to each individual eclipsing binary distance with respect to the centre of this galaxy. As expected, all corrections (listed in Table 1) are very small. Applying them we calculated the same mean distance of 18.476 ± 0.002 mag. However, our sample of systems is large enough to independently compute a geometric model of the LMC inner disk. We adopted the centre of the young stellar population in the LMC (right ascension RA = 5 h 20 min 12 s, declination dec. = –69° 18′ 00″ J2000.0) 15 as the centre of this galaxy, and using the least squares method fitted a plane to our data. We obtained an inclination angle of 25° ± 4°, a positional angle of 132° ± 10°, and a mean distance modulus of 18.477 ± 0.004 mag, with a reduced $\chi^2$ very close to unity. This determination of the distance error is based on Monte Carlo simulations (see Extended Data Fig. 4), so it is larger than the statistical errors on the mean distances reported before. Our geometrical parameters of the LMC inner disk agree very well with the corresponding parameters obtained by other authors 14–16 . We have thus demonstrated that the geometry of the LMC does not affect our distance determination, within the quoted error.

The systematic uncertainty on the LMC distance includes the following contributions: calibration of the surface brightness–colour relation (0.018 mag, or 0.8%), photometric zero points (0.01 mag for both V and K bands, or 0.5%), and reddening absolute scale (0.013 mag, or 0.6%). Combining these quadratically we obtain 0.026 mag, or 1.1%.

Our distance determination might in principle depend on the metallicity and gravity via a sensitivity of the surface brightness–colour relation to these parameters. However, the relation depends only very weakly on gravity; it has been demonstrated 6 that the difference between relations obtained from giant and dwarf stars is about 1%. Since our LMC binary stars and nearby clump stars are all giants we do not expect any contribution to the total error from gravity differences. The relation is also virtually independent of metallicity 17 . Our sample of local calibrating red clump giant stars spans a large range of metallicities (1 dex). We studied the residuals between observed and calculated values of the relation (see Extended Data Fig. 2) and showed, in agree-ment with other studies 17 , that there is no dependence on metallicity (see Methods for more details). Moreover, the range of metallicities of the nearby stars used for the calibration of the relation overlaps with the metallicities of our eclipsing binaries in the LMC.

We therefore adopt 18.477 ± 0.004 (statistical) ± 0.026 (systematic) mag as our final, best value for the LMC distance modulus.

The LMC is a unique 'laboratory' for very detailed studies of many different objects and processes. Most of these studies require the knowledge of a precise distance to the galaxy. This requirement will become more and more important with the advent of new giant telescopes that will enable a wide variety of studies of the LMC and other nearby galaxies with an unprecedented precision. The LMC hosts large samples of different

'standard candles' including classical Cepheids, and therefore plays a key part in the calibration of the extragalactic distance scale and determination of the Hubble constant[6]. For this reason, there is an impressive list of more than 600 distance determinations of the LMC in the literature[18]. Unfortunately, very few of them provide even a crude estimation of the systematic errors affecting these measurements, which indicates how difficult it is to determine a precise distance to even the closest galaxy.

In about 5 years, we expect that Galactic Cepheid period–luminosity (P–L) relations will be calibrated from the individual distances of Cepheids inferred from their parallaxes as obtained by the Gaia mission. The overall precision is expected to be similar to that of the corresponding LMC Cepheid P–L relations 11,19 , when calibrated with our LMC distance. An important advantage of the LMC relations is their weaker dependence on the reddening as compared to the P–L relations derived from the more strongly reddened Galactic Cepheids. Most probably the best way to obtain the final absolute Cepheid calibration will be to average the two independent zero points.

As is evident from Fig. 3, eclipsing binaries offer us an opportunity to determine stellar distances that are already competitive in precision to expected Gaia-derived distances at about 1 kpc from the Sun, and that retain this precision as far as the outskirts of the Local Group of galaxies: with the advent of the new extremely large telescopes in the near future, such precision could be extended to galaxies far beyond the Local Group.

Distances determined with the binary technique will continue to be very important for three reasons. First, they will provide a check on Gaia parallaxes at a 1% precision level. Second, precise distances to many nearby galaxies can and will soon be determined with this technique. Last, the method provides an independent precision zero point for the extragalactic distance scale and the calibration of the Cepheid P–L relation in different environments, paving the way for a Hubble constant determination precise to 1% from the combined Cepheid P–L relation–supernova Ia method.

We expect that the accuracy of our method of distance determination will soon be verified with nearby Galactic eclipsing binary systems for which geometrical distances will become available from Gaia data, and in some cases also with nearby astrometric binaries. A comparison of two independent geometric distance determinations for one such system (TZ Fornacis) shows excellent agreement (see Methods). In the future, with the final Gaia parallaxes available, such a test will be possible using many eclipsing binaries.

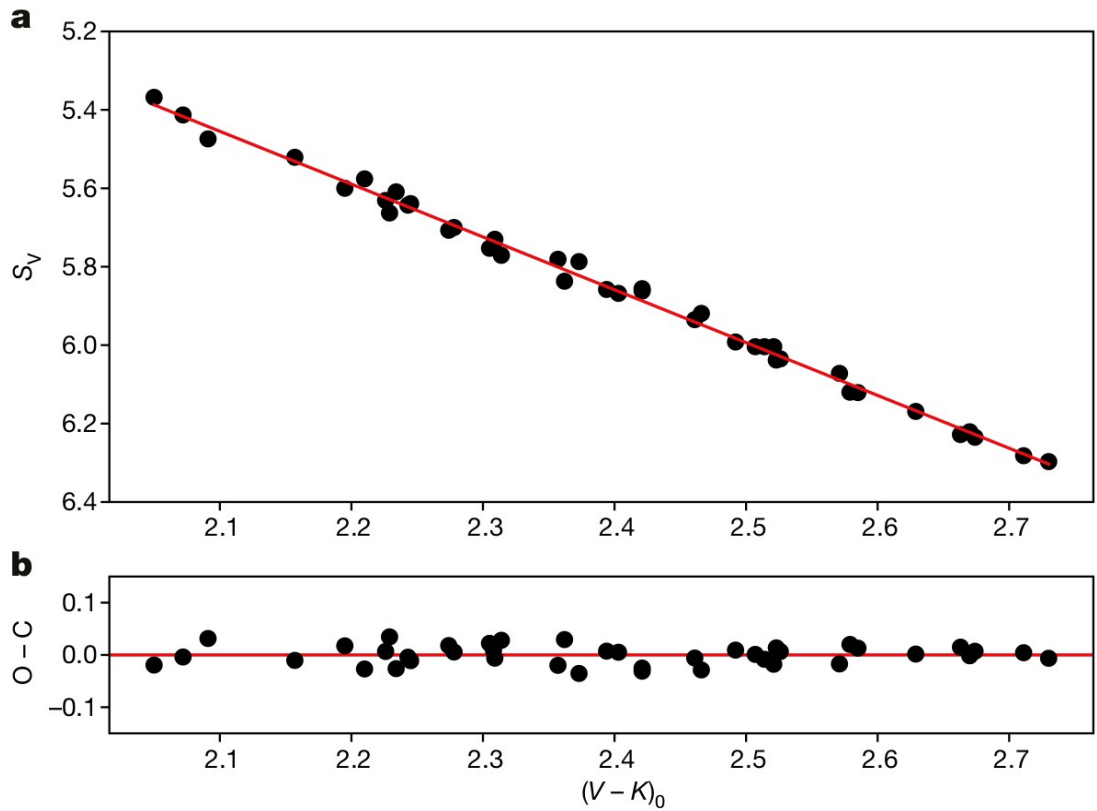

**Fig. 1. New relation between surface brightness $S_V$ and $(V - K)_0$ colour.** a, Plot of $S_V$ versus $(V - K)_0$ (data points) and fitted line. The r.m.s. scatter on this relation is 0.018 mag, which means an 0.8% precision in stellar angular diameters. The precision of this relationship is fundamental for measuring precise cosmic distances. b, Residuals (observed minus calculated) for the fit. Both $S_V$ and $(V - K)_0$ are in magnitudes.

| System | RA | DEC | (m-M) | $\sigma_{m-M}$ | corr |
|---|---|---|---|---|---|
| OGLE-LMC- | (h min s) | (° ′ ″) | (mag) | (mag) | (mag) |
| ECL-01866 | 04:52:15.28 | -68:19:10.30 | 18.515 | 0.031 | -0.028 |
| ECL-03160 | 04:55:51.48 | -69:13:48.00 | 18.474 | 0.013 | -0.038 |
| ECL-05430 | 05:01:51.74 | -69:12:48.80 | 18.522 | 0.012 | -0.028 |
| ECL-06575 | 05:04:32.87 | -69:20:51.00 | 18.483 | 0.011 | -0.026 |
| ECL-09114 | 05:10:19.64 | -68:58:12.20 | 18.490 | 0.028 | -0.009 |
| ECL-09660 | 05:11:49.45 | -67:05:45.20 | 18.465 | 0.019 | 0.029 |
| ECL-09678 | 05:11:51.76 | -69:31:01.10 | 18.501 | 0.018 | -0.017 |
| ECL-10567 | 05:14:01.89 | -68:41:18.20 | 18.455 | 0.014 | 0.002 |
| SC9-230659 | 05:14:06.04 | -69:15:56.90 | 18.456 | 0.026 | 0.009 |
| ECL-12669 | 05:19:12.80 | -69:06:44.40 | 18.450 | 0.019 | -0.002 |
| ECL-12875 | 05:19:45.39 | -69:44:38.50 | 18.453 | 0.026 | -0.009 |
| ECL-12933 | 05:19:53.69 | -69:17:20.40 | 18.476 | 0.025 | 0.000 |
| ECL-13360 | 05:20:59.46 | -70:07:35.20 | 18.489 | 0.013 | -0.004 |
| ECL-13529 | 05:21:23.34 | -70:33:00.00 | 18.498 | 0.016 | 0.005 |
| ECL-15260 | 05:25:25.66 | -69:33:40.50 | 18.453 | 0.034 | 0.003 |
| ECL-18365 | 05:31:49.56 | -71:13:28.30 | 18.479 | 0.021 | 0.033 |
| ECL-18836 | 05:32:53.06 | -68:59:12.30 | 18.473 | 0.018 | 0.026 |
| ECL-21873 | 05:39:51.19 | -67:53:00.50 | 18.445 | 0.014 | 0.059 |
| ECL-24887 | 05:50:39.02 | -69:14:20.70 | 18.515 | 0.023 | 0.045 |
| ECL-25658 | 06:01:58.77 | -68:30:55.10 | 18.423 | 0.016 | 0.076 |

**Table 1. Distance moduli of the studied eclipsing binary systems.** The first, second and third columns give the name, right ascension and declination, respectively, of the systems studied. The fourth column gives the distance modulus. The fifth column gives the total statistical uncertainty for the mean distance modulus estimated on the basis of Monte Carlo simulations. The geometrical corrections calculated from the model[14] are given in the last column.

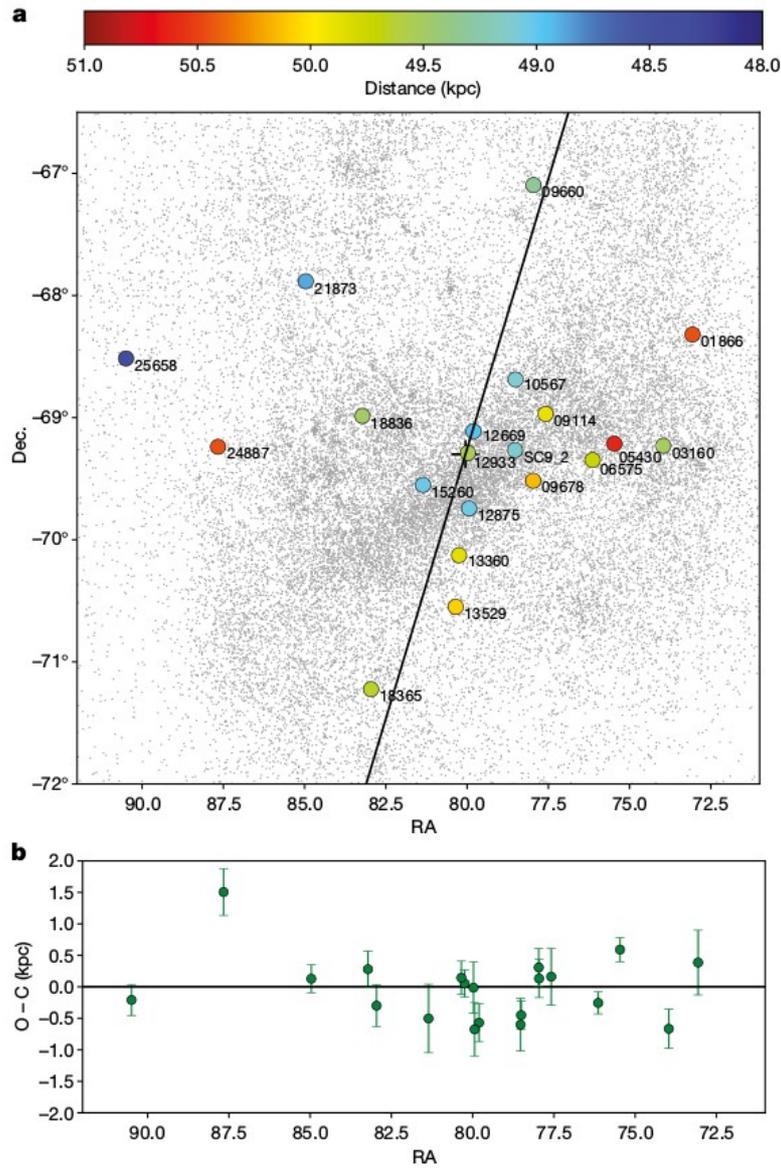

**Fig. 2. Locations and distances of our 20 eclipsing binary systems in the LMC.** a, The distribution of the observed systems over the map of the central regions of the LMC. The line of node is shown with a solid line. The full names of the systems are given in Table 1. Different colour codes denote the distances to individual systems (see colour key). b, Distance offsets between individual systems and the best fitted LMC disk plane, plotted versus right ascension (in degrees). The error bars correspond to $1\sigma$ errors.

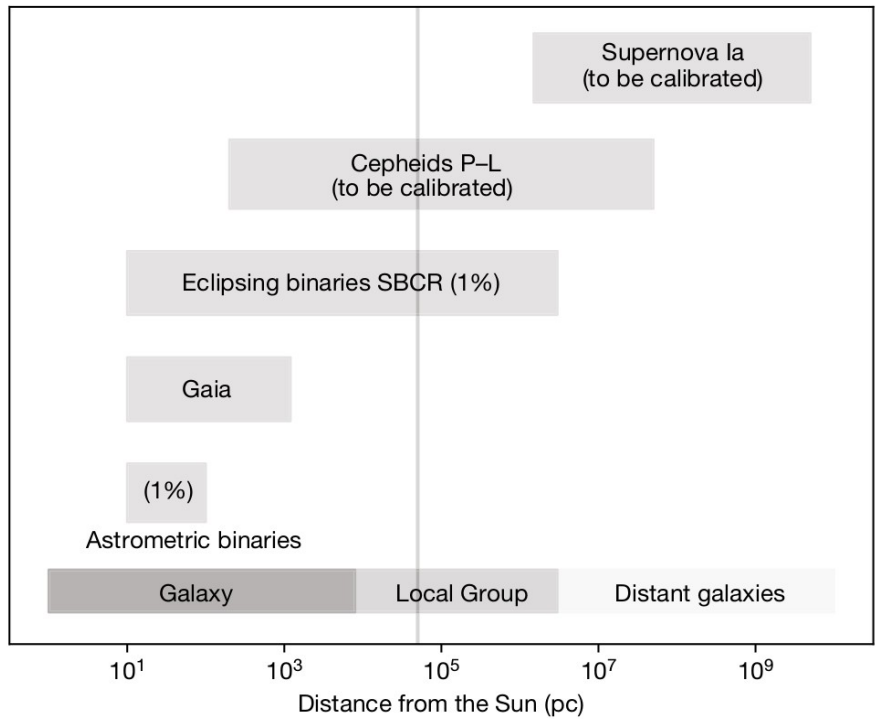

**Fig 3. Range and precision of the geometrical methods.** The range of distances that can be measured with astrometric binaries, Gaia parallaxes, eclipsing binaries together with the surface brightness–colour relation (SBCR), the Cepheid P–L relation and type Ia supernovae are marked with the corresponding bars. Astrometric binaries, Gaia parallaxes, and eclipsing binaries with the SBCR can be used to calibrate the Cepheid P–L relation, and as a result the supernova Ia brightness. As can be appreciated, eclipsing binaries together with our new relation offer us the opportunity to determine distances competitive in precision to Gaia already at about 1 kpc from the Sun, while retaining their high precision for distances up to 1 Mpc. The lower bar divided into three sections shows the range of distances within our Galaxy, Local Group, and more distant galaxies. The vertical grey line marks the distance to the LMC.

## Methods

**Galactic red clump stars.**
*Interferometric data.* Observations were performed using the PIONIER instrument of the Very Large Interferometer array (VLTI). PIONIER combines the light coming from four telescopes in the H band, either in a broad band mode or with a low spectral resolution, where the light is dispersed across a few spectral channels. Observations, data reductions and angular diameter estimates of our stars are detailed elsewhere [8]. Briefly, squared visibilities and closure phases are extracted from the interference fringe pattern for each telescope pair and triplet and for each spectral channel. These main interferometric observables are then calibrated with the observations of a calibrator star in order to derive the transfer function. Angular diameters are obtained by fitting the calibrated squared visibilities with the analytical formulae of the visibility function. Half of the angular diameters determined have formal errors better than 1.2%, while the overall accuracy is better than 2.7%[8].

*Photometric data.* We adopted V band magnitudes from the General Catalogue of Photometric Data [9]. This catalogue contains measurements from different sources homogenized and transformed onto the Johnson system. For many stars from our sample we have multiple (5–7) measurements. The mean scatter between these measurements is on average 0.008 mag, which demonstrates overall excellent agreement of the zero points from different sources. The compilation of individual V band photometry for our stars is presented in Extended Data Table 1. Conservatively, we decided to adopt the mean V band magnitudes as they are in the General Catalogue of Photometric Data and do not introduce any additional subjective cleaning of individual measurements.

In order to perform an external check we transformed photometry of our stars from the Tycho catalogue[20] onto the Johnson system. We checked two different calibrations[20,21] and the mean difference between them was −0.002 mag, which shows that the calibration does not introduce any significant error. Then we compared the transformed V band Tycho photometry with our adopted V band photometry[9] and we found a difference of 0.002 mag only.

The high precision JHK band photometry for our nearby red clump stars[7] was obtained at the South African Astronomical Observatory (SAAO), with the 0.75-m telescope and the Mk II infrared photometer. This equipment defines the SAAO JHLK standard system and has been used for 30 years to obtain precise JHK measurements of bright stars[22]. The mean standard error in magnitude is close to 0.01 mag including all systematics. The uncertainty of the individual measurements is dominated by the quality of standards (about 0.006 mag)[23]. Based on stars with multiple observations, the internal standard deviation of a single measurement is 0.005 mag in the K band. Since a given star was observed at different positions on the sky and was transformed to the standard system using different standards this error contains most of the error of standardization. The error of standardization was estimated to be about 0.004 mag[7].

To check on variability of our sample stars we analysed in detail their Hipparcos V band light curves[20]. These light curves typically contain 140 measurements. We calculated average standard deviation of 0.0068 mag with the largest being 0.009 mag. This together with the comparison of multiple observations in optical and near infrared bands clearly shows that these stars do not change their magnitudes by more than 0.007 mag. Such photometric behaviour is expected for red clump stars in a quiet evolutionary phase of the core helium burning.

*Extinction correction.* Interstellar extinction E(B − V) was estimated using the reddening maps[24,25]. The maps give a total extinction in a given direction, while the reddening to a red clump star in our sample is only a small fraction of it. To calculate this fraction we used a simple exponential model of dust distribution in the Milky Way, the Hipparcos parallaxes and the local, extinction free, bubble radius of 40 pc [26]. Such a procedure works very well for nearby stars.

In order to further check the calculated reddenings we used two additional methods. First, we adopted effective temperatures[8] and based on three different calibrations calculated the intrinsic colours[27–29]. Final intrinsic colours for our stars were adopted as a mean from the three determinations. Comparing them to the observed colours we calculated the reddenings. They are listed in the second column of Extended Data Table 3. In spite of a relatively large scatter due to the low precision of this method, the obtained results are in a good agreement with the reddening values from the Schlegel maps[24].

The second method is based on fitting the spectral energy distributions (SED). The synthetic photospheric emission for our stars was modelled with tabulated stellar atmosphere models from the ATLAS9 code[30]. We chose a grid that was computed for solar metallicity and a standard turbulence velocity of 2 km s$^{-1}$. The grid was then interpolated to compute spectra for any effective temperature and any surface gravity. The spectrum is also multiplied by the solid angle of the stellar photosphere corresponding to the limb-darkened angular diameter. We adjusted the photometric data to the model taking into account the photometric band-passes and zero-points of each filter. We fitted E(B − V) and $T_{eff}$ from the SED with photometry from Gaia (G, Bp, Rp)[31], Tycho-2 (B, V)[20], 2MASS (JHK)[7], and Wise (<5 μm only)[32]. The mean difference between these reddenings and the reddenings based on Schegel's map is 0.007 ± 0.01 mag.

We adopted the reddenings obtained from Schlegel's maps[24] in our work. We note that the obtained reddenings are very small (typically E(B − V) is much smaller than 0.01 mag). Therefore, the correction for extinction should not affect our result of the distance determination to the LMC. To check it in more detail, we calculated surface brightness–colour relations assuming a uniform zero E(B − V) reddening for our red clump stars, 0.007 mag, and a random uniform distribution of values from 0 to 0.018 mag (the largest reddening in our sample). Then we calculated mean distances to the LMC based on these two relationships and compared them to our best-fit value. We obtained the following changes in distance: 0.14%, 0.05% and 0.07%, respectively, demonstrating that the very small correction for extinction has a negligible effect on our LMC distance measurement.

*Determination of the surface brightness–colour relation and comparison to other studies.* Our final surface brightness–colour relation was obtained from a linear least squares fit using errors on $S_V$ only (that is, along the y axis). We adopted errors on V band magnitudes of 0.007 mag, and the errors on the angular diameters listed in Extended Data Table 2. We used a mean-centred relation in order to have an uncorrelated parametrization of the slope and zero point, enabling the zero-point to be much better constrained[33]. The optimal value of $x_0 = 2.405$ is determined according to our data set.

We performed a bootstrapping analysis with replacement with 1,000 bootstrap samples and we obtained the following relation:

$$S_V = (1.330 \pm 0.017) \times [(V-K)_0 - 2.405] + (5.869 \pm 0.003)$$

$$\text{r.m.s. scatter} = (0.018 \pm 0.002) \text{ mag}$$

The uncertainty is estimated using the bootstrapped interval between the 16th and 84th percentiles of the distribution between the median and maximum value.

We also used several different versions of least squares fitting: (1) without taking into account the observational errors; (2) assuming errors on both axes with a uniform error bar of 0.005 mag on the K band photometry; and (3) assuming errors on the V and K band magnitudes of 0.01 mag. In all cases we obtained very similar results. The obtained $S_V$ values differ at most by 0.002 mag, and the errors on the zero point are in the range from 0.003 to 0.004 mag. The r.m.s scatter in all cases remains the same. We thus demonstrated that our distance determination to the LMC in insensitive to a change of weighting of the data and computation method of the fit to a level of 0.05%.

**Distance determination.** Precise astrophysical parameters of all our eclipsing binary systems were already obtained 10,11 . The individual distances were calculated in the same manner as in our previous papers[5,34]. There are no statistically significant correlations between the distances to individual components and the effective temperatures or gravities of the components.

The accuracy of distance determinations based on our surface brightness–colour relation can be verified with Galactic eclipsing binaries. Indeed, independent, very precise distances to many Galactic eclipsing binaries will be provided by the Gaia mission.

Having both astrometric and spectroscopic orbits of binary systems, one can also obtain direct geometric distances with a precision better than 1%[35,36]. However, the systems need to be spatially resolved to enable astrometric measurements, which limits the range of this method (see Fig. 3). Therefore eclipsing binary systems that are located close enough to obtain an astrometric orbit for them are also very well suited to check on the accuracy of our approach.

TZ For is a nearby eclipsing binary system composed of a giant (primary) and subgiant (secondary) star. Precise physical parameters of this system were obtained from an analysis of spectroscopic and photometric data[37]. Based on these parameters and applying our new surface brightness–colour relation, we obtained the following distance to this system: 185.1 ± 2.0 (stat.) ± 1.9 (syst.) pc. The statistical error is dominated by error on the radius determination of the giant. A very precise distance to TZ For of 186.1 ± 1.0 (stat. + syst.) pc was obtained based on spectroscopic and astrometric orbits[36,38]. The two determinations are in excellent agreement. The distance to TZ For from its Gaia DR2 parallax is 183.4 ± 0.8 (stat.) pc. However several independent studies have revealed that there is a zero-point shift of Gaia DR2 parallaxes of about 0.03 mas[39–41]. Therefore we assume a systematic uncertainty on the current DR2 Gaia distance of 1.3 pc, which brings all three independent geometrical distance determinations of TZ For into very good agreement.

With the final Gaia parallaxes expected in 5 years from now, a comparison between Gaia distances and distances determined for binaries with our surface brightness–colour relation will be performed for many eclipsing binaries at the 1% level. This future work will allow us to definitively test and verify the accuracy of our method.

**Code availability**

We do not provide any code because we used only classical tools such as the IRAF, Daophot and Wilson–Devinney code, and they are publicly available.

**Data availability**

All data are available upon request from G.P.

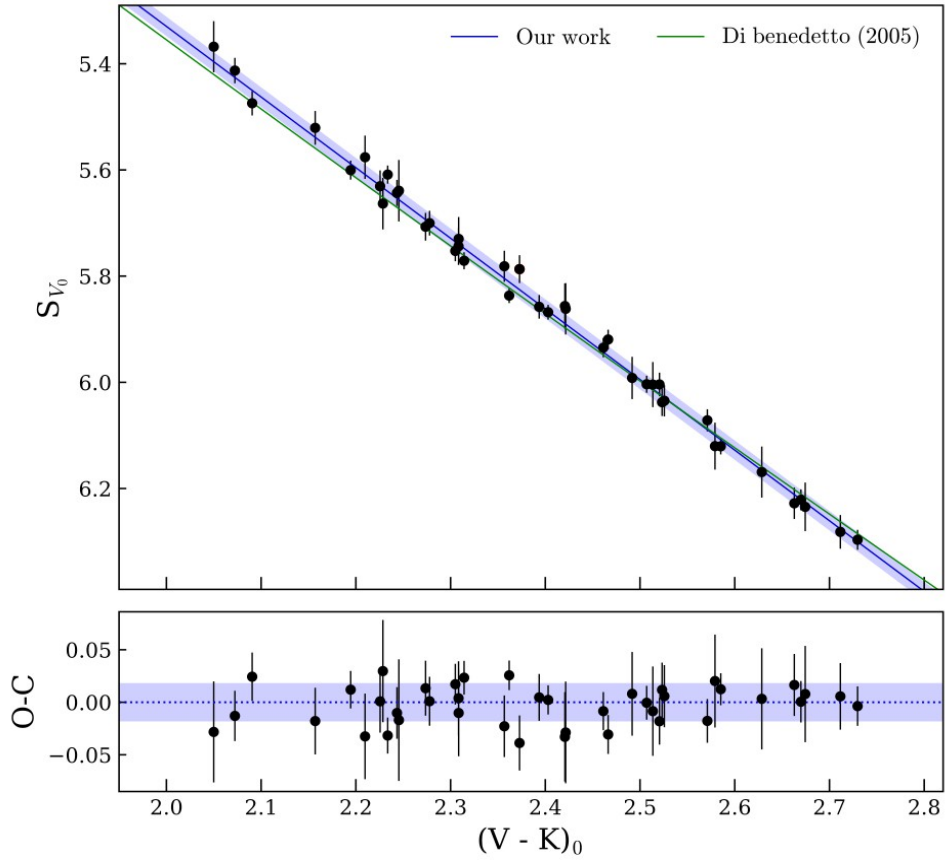

**Extended Data Fig.1. Comparison of our relation with the relation of Di Benedetto obtained for giant stars[6].** Top panel, comparison of relations: data points show our results, with the fitted line shown in blue. The blue shaded area represents our obtained r.m.s. scatter of 0.018 mag. The green line is from ref.[6]. Very good agreement is demonstrated. Both $S_V$ and $(V - K)_0$ are in magnitudes. $S_V$ physically corresponds to the V band magnitude of a red giant star whose angular diameter is 1 mas. The error bars correspond to $1\sigma$ errors. Bottom panel, observed minus calculated values.

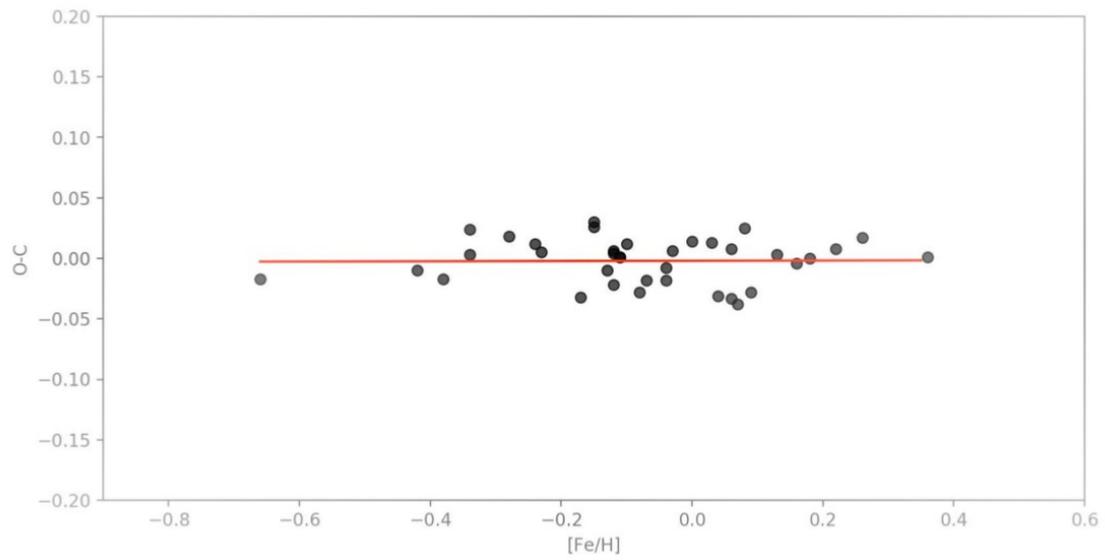

**Extended Data Fig.2. Observed minus calculated surface brightness versus metallicity[6], [Fe/H].** In a relatively large range of metallicities (about 1 dex) no correlation is found. A formal linear fit gives O − C = 0.0009[Fe/H] – 0.002 dex with coefficient of determination $R^2$ = 0.0001.

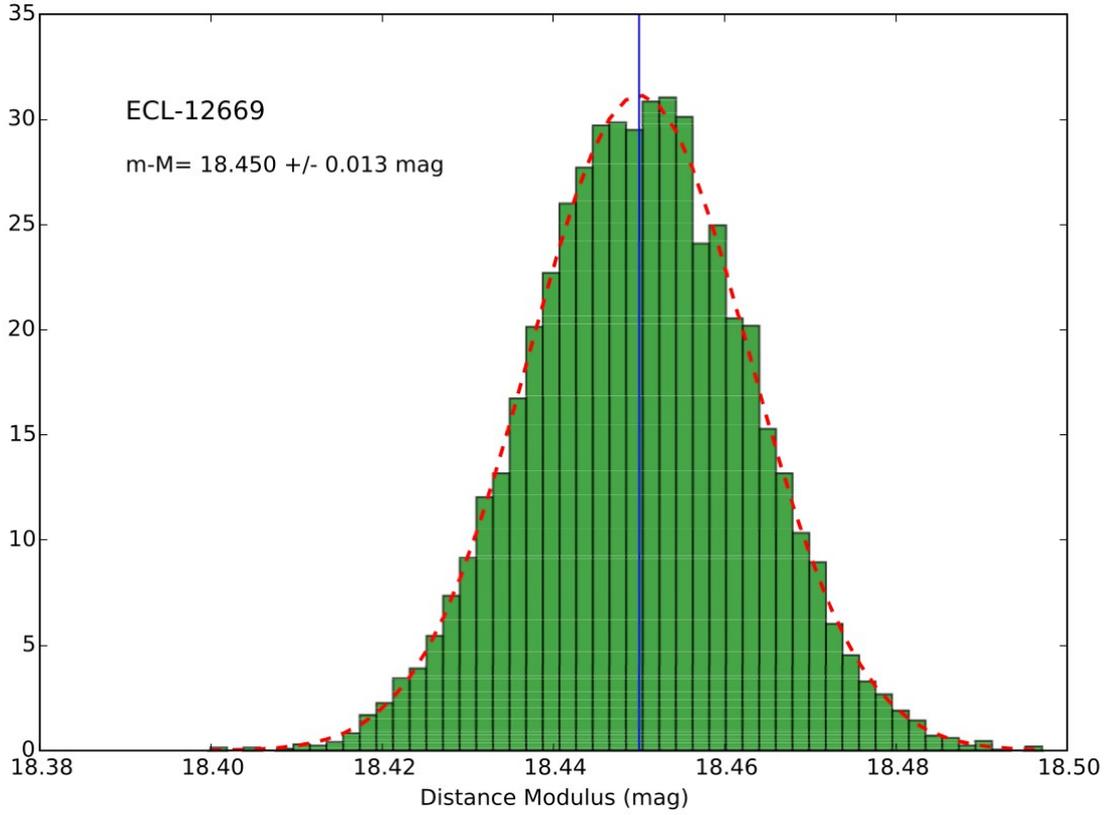

**Extended Data Fig.3. Example of Monte Carlo simulations for one of our objects, ECL-12669.** We computed 10,000 models with the JKTEBOP code[77] from which we obtained statistical uncertainties on the radii $R_1$ and $R_2$, the orbital inclination i, the phase shift φ, the surface brightness ratio $j_{21}$, radial velocity semi-amplitudes $K_1$ and $K_2$, and the systemic velocities $γ_1$ and $γ_2$. For every model we computed the distance modulus converting $j_{21}$ into temperature ratio $T_2/T_1$ by using Popper's calibration[78] and our original solution with the Wilson–Devinney code[79]. We plot the number of calculated models versus distance modulus *(m − M)*. The dashed line is the best fitted Gaussian and the blue line is the distance determined for this object. The intrinsic $(V − K)_0$ colours used to estimate the angular diameters of the components were computed using a temperature–colour calibration[28].

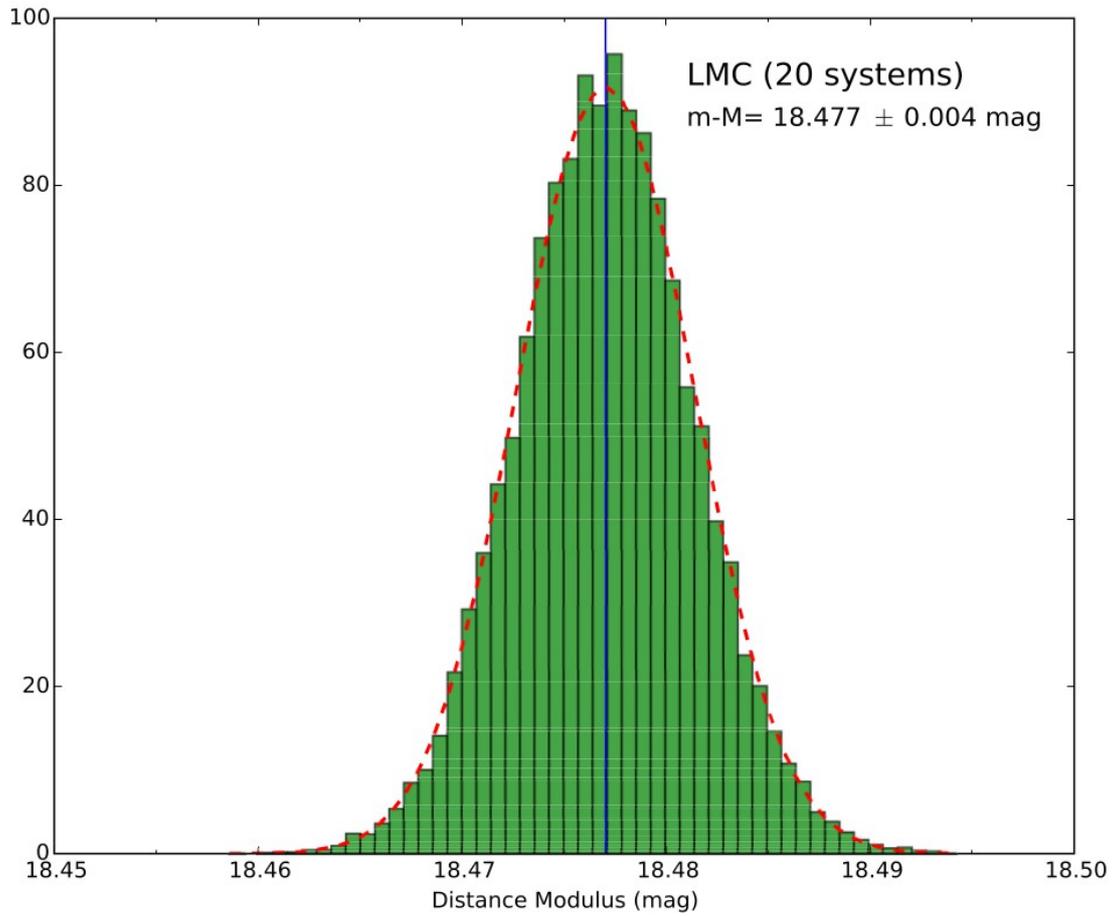

Extended Data Fig.4. Error estimate of distance modulus of the LMC from Monte Carlo simulations. For each eclipsing binary we calculated random 20,000 distance moduli with Gaussian distribution assuming $\mu = m - M_{mean}$ and $\sigma = \sigma_{mean}$ from Table 1. Then we fitted 20,000 planes with the linear least-squares method for every set of distance moduli using as free parameters inclination of the disk plane, $i$, the position of the nodes, $\Theta$, and the distance to the centre of the LMC, $d$. Apparent positions were converted into the three-dimensional Cartesian positions[80]. We plot the number of calculated models versus distance modulus $(m - M)$. The dashed line is the best fitted Gaussian and the blue line is the distance of the LMC.

**Extended Data Table 1. V band magnitudes of our target stars**

| Star HD | V band measurements [mag] | References |
|---|---|---|
| 360 | 5.99 5.98 5.99 5.98 | 20 21 24 29 |
| 3750 | 6.01 6.00 6.01 6.00 | 20 30 31 45 |
| 4211 | 5.904 5.90 5.85 5.83 | 42 47 53 |
| 5722 | 5.615 5.61 5.64 | 49 53 |
| 8651 | 5.42 5.42 5.422 5.423 5.41 5.40 5.41 5.42 | 20 21 25 27 30 33 39 51 |
| 9362 | 3.95 3.944 3.94 3.94 3.91 3.94 3.95 3.94 | 20 25 28 30 31 39 51 52 |
| 10142 | 5.94 5.93 5.95 | 20 44 53 |
| 11977 | 4.69 4.68 4.68 4.71 | 20 30 50 52 |
| 12438 | 5.35 5.34 5.34 5.360 | 20 33 40 54 |
| 13468 | 5.93 5.92 5.95 | 20 32 53 |
| 15220 | 5.88 5.87 5.88 5.89 | 20 30 37 53 |
| 15248 | 6.01 6.00 6.00 | 20 30 50 |
| 15779 | 5.35 5.34 5.37 | 20 32 53 |
| 16815 | 4.11 4.10 4.11 4.11 | 20 30 40 51 |
| 17652 | 4.46 4.45 4.45 4.46 | 20 30 40 51 |
| 17824 | 4.74 4.76 4.76 4.77 4.78 | 20 30 40 51 53 |
| 18784 | 5.75 5.75 5.74 5.75 | 20 30 35 53 |
| 23319 | 4.59 4.58 4.58 | 20 30 41 |
| 23526 | 5.91 5.90 5.92 | 20 31 34 |
| 23940 | 5.551 5.54 5.53 5.54 | 20 30 47 |
| 30814 | 5.05 5.03 | 49 53 |
| 36874 | 5.76 5.78 | 49 53 |
| 39523 | 4.51 4.50 4.50 4.49 | 30 40 53 |
| 39640 | 5.17 5.16 5.16 | 20 30 41 |

| HD | V magnitudes | Refs |
|---|---|---|
| 39910 | 5.87 5.86 5.86 | 20 23 30 |
| 40020 | 5.874 5.88 | 53 |
| 43899 | 5.53 5.55 5.58 | 33 38 53 |
| 46116 | 5.38 5.37 5.37 5.37 5.38 | 20 30 40 50 53 |
| 53629 | 6.092 6.08 | 38 |
| 56160 | 5.58 5.58 | 49 53 |
| 60060 | 5.874 5.87 | 38 |
| 60341 | 5.66 5.63 | 46 53 |
| 62713 | 5.17 5.12 5.16 5.12 5.16 | 20 26 30 36 53 |
| 68312 | 5.36 5.34 5.36 5.37 | 20 35 44 53 |
| 74622 | 6.29 6.29 6.28 6.26 6.28 | 20 30 41 45 53 |
| 75916 | 6.128 6.11 | 38 |
| 176704 | 5.65 5.64 5.65 | 20 48 53 |
| 191584 | 6.22 6.21 6.20 6.24 | 20 30 36 53 |
| 204381 | 4.51 4.49 4.49 4.51 4.50 4.510 | 20 30 40 51 52 54 |
| 219784 | 4.416 4.41 4.40 4.40 4.43 4.41 4.420 | 20 30 40 43 51 54 |
| 220572 | 5.588 5.60 5.65 | 37 53 |

These data come from the General Catalog of Photometric data[9]. Left column, HD number; middle column, V band magnitude; right column, refs.[42–74], see page 352 of ref.[9] and ref.[75], and ref.[76]; also T. Oja, personal communication, and L. Haggkvist & T. Oja, personal communication.

**Extended Data Table 2. Collected data for our sample of helium burning giants**

| Star | LD [mas] | $\sigma$LD [mas] | V [mag] | $\sigma$V [mag] | K [mag] | E(B-V) [mag] | Sv [mag] | $\sigma$Sv [mag] |
|---|---|---|---|---|---|---|---|---|
| HD360  | 0.906 | 0.014 | 5.986 | 0.005 | 3.653 | 0.009 | 5.744 | 0.036 |
| HD3750 | 1.003 | 0.019 | 6.004 | 0.005 | 3.485 | 0.002 | 6.004 | 0.043 |
| HD4211 | 1.100 | 0.009 | 5.877 | 0.028 | 3.295 | 0.004 | 6.072 | 0.035 |
| HD5722 | 0.995 | 0.018 | 5.618 | 0.012 | 3.381 | 0.010 | 5.576 | 0.042 |
| HD8651 | 1.228 | 0.011 | 5.418 | 0.006 | 3.019 | 0.002 | 5.858 | 0.023 |
| HD9362 | 2.301 | 0.017 | 3.943 | 0.006 | 1.638 | 0.000 | 5.753 | 0.020 |

| Star | LD | σ_LD | V | σ_V | K | E(B−V) | SV | σ_SV |
|---|---|---|---|---|---|---|---|---|
| HD10142 | 0.964 | 0.004 | 5.938 | 0.009 | 3.557 | 0.007 | 5.837 | 0.017 |
| HD11977 | 1.528 | 0.010 | 4.686 | 0.011 | 2.486 | 0.002 | 5.600 | 0.021 |
| HD12438 | 1.091 | 0.015 | 5.344 | 0.007 | 3.176 | 0.004 | 5.521 | 0.033 |
| HD13468 | 0.886 | 0.009 | 5.934 | 0.013 | 3.666 | 0.009 | 5.643 | 0.028 |
| HD15220 | 1.185 | 0.015 | 5.881 | 0.006 | 3.199 | 0.007 | 6.228 | 0.030 |
| HD15248 | 0.949 | 0.018 | 6.001 | 0.003 | 3.553 | 0.010 | 5.856 | 0.043 |
| HD15779 | 1.185 | 0.013 | 5.357 | 0.013 | 3.067 | 0.006 | 5.707 | 0.029 |
| HD16815 | 2.248 | 0.009 | 4.109 | 0.003 | 1.706 | 0.000 | 5.868 | 0.014 |
| HD17652 | 1.835 | 0.010 | 4.456 | 0.005 | 2.139 | 0.001 | 5.771 | 0.017 |
| HD17824 | 1.391 | 0.013 | 4.764 | 0.010 | 2.668 | 0.002 | 5.474 | 0.025 |
| HD18784 | 1.036 | 0.013 | 5.748 | 0.004 | 3.353 | 0.014 | 5.781 | 0.030 |
| HD23319 | 2.033 | 0.010 | 4.583 | 0.005 | 1.995 | 0.001 | 6.121 | 0.016 |
| HD23526 | 0.915 | 0.020 | 5.909 | 0.008 | 3.634 | 0.017 | 5.663 | 0.049 |
| HD23940 | 1.093 | 0.020 | 5.543 | 0.007 | 3.229 | 0.002 | 5.730 | 0.042 |
| HD30814 | 1.310 | 0.008 | 5.041 | 0.010 | 2.791 | 0.006 | 5.609 | 0.020 |
| HD36874 | 1.118 | 0.010 | 5.768 | 0.010 | 3.242 | 0.002 | 6.004 | 0.024 |
| HD39523 | 1.939 | 0.013 | 4.500 | 0.006 | 2.036 | 0.001 | 5.935 | 0.019 |
| HD39640 | 1.251 | 0.016 | 5.163 | 0.005 | 2.921 | 0.006 | 5.631 | 0.030 |
| HD39910 | 1.090 | 0.006 | 5.863 | 0.005 | 3.315 | 0.015 | 6.004 | 0.017 |
| HD40020 | 1.012 | 0.022 | 5.876 | 0.003 | 3.419 | 0.013 | 5.862 | 0.049 |
| HD43899 | 1.264 | 0.016 | 5.557 | 0.015 | 3.004 | 0.010 | 6.035 | 0.033 |
| HD46116 | 1.145 | 0.030 | 5.373 | 0.005 | 3.103 | 0.009 | 5.639 | 0.058 |
| HD53629 | 1.065 | 0.023 | 6.085 | 0.006 | 3.410 | 0.017 | 6.169 | 0.049 |
| HD56160 | 1.411 | 0.010 | 5.580 | 0.000 | 2.823 | 0.010 | 6.297 | 0.019 |
| HD60060 | 0.948 | 0.009 | 5.872 | 0.002 | 3.545 | 0.018 | 5.700 | 0.023 |
| HD60341 | 1.190 | 0.021 | 5.645 | 0.015 | 3.126 | 0.010 | 5.992 | 0.043 |
| HD62713 | 1.446 | 0.010 | 5.134 | 0.017 | 2.654 | 0.005 | 5.919 | 0.025 |
| HD68312 | 1.020 | 0.022 | 5.359 | 0.008 | 3.279 | 0.011 | 5.368 | 0.049 |
| HD74622 | 1.020 | 0.014 | 6.279 | 0.010 | 3.532 | 0.013 | 6.282 | 0.033 |
| HD75916 | 1.013 | 0.020 | 6.117 | 0.009 | 3.516 | 0.008 | 6.120 | 0.045 |
| HD176704 | 1.317 | 0.010 | 5.645 | 0.005 | 2.956 | 0.007 | 6.221 | 0.020 |
| HD191584 | 1.024 | 0.021 | 6.211 | 0.012 | 3.512 | 0.009 | 6.235 | 0.047 |
| HD219784 | 2.117 | 0.023 | 4.412 | 0.008 | 1.886 | 0.001 | 6.038 | 0.025 |
| HD220572 | 1.092 | 0.012 | 5.605 | 0.020 | 3.224 | 0.003 | 5.787 | 0.027 |
| HD204381 | 1.524 | 0.015 | 4.501 | 0.008 | 2.426 | 0.001 | 5.413 | 0.033 |

We present measured angular diameters[8] (LD), their errors ($\sigma_{LD}$), V[9] magnitudes and the corresponding errors $\sigma_V$, K[7] magnitudes, estimated reddening $E(B − V)$, and calculated surface brightness SV and the corresponding errors ($\sigma_{SV}$). The errors on V band magnitudes are from ref.[9] and were calculated from the dispersion from multiple measurements (typically 4–6). The typical dispersion of the V band magnitudes measured from the scatter of their Hipparcos V band photometry is 0.007 mag and we adopted this value as the statistical error on the V band measurements. The typical precision of the K band photometry is about 0.005 mag.

**Extended Data Table 3. Reddenings determined with three different methods**

| Star | E(B-V) Schlegel | E(B-V)$_1$ Colors | E(B-V)$_2$ SED |
|---|---|---|---|
| HD360 | 0.009 | 0.002 | 0.013 |
| HD3750 | 0.002 | 0.010 | 0.009 |
| HD4211 | 0.004 | 0.009 | 0.011 |
| HD5722 | 0.010 | -0.009 | 0.008 |
| HD8651 | 0.002 | 0.005 | 0.012 |
| HD9362 | 0.000 | -0.011 | 0.010 |
| HD10142 | 0.007 | 0.003 | 0.015 |
| HD11977 | 0.002 | -0.000 | 0.010 |
| HD12438 | 0.004 | -0.013 | 0.001 |
| HD13468 | 0.009 | 0.002 | 0.009 |
| HD15220 | 0.007 | 0.044 | 0.030 |
| HD15248 | 0.010 | 0.015 | 0.011 |
| HD15779 | 0.006 | 0.003 | 0.015 |
| HD16815 | 0.000 | -0.007 | 0.009 |
| HD17652 | 0.001 | -0.008 | 0.020 |
| HD17824 | 0.002 | -0.017 | 0.018 |
| HD18784 | 0.014 | 0.001 | 0.009 |
| HD23319 | 0.001 | 0.052 | 0.033 |
| HD23526 | 0.017 | 0.009 | 0.016 |
| HD23940 | 0.002 | -0.004 | 0.003 |
| HD26464 | 0.008 | 0.027 | 0.021 |
| HD30814 | 0.006 | 0.008 | 0.016 |
| HD35369 | 0.000 | -0.001 | 0.020 |
| HD36874 | 0.002 | 0.007 | 0.009 |
| HD39523 | 0.001 | 0.016 | 0.016 |
| HD39640 | 0.006 | 0.005 | 0.013 |
| HD39910 | 0.015 | 0.021 | 0.022 |
| HD40020 | 0.013 | 0.018 | 0.013 |
| HD43899 | 0.010 | 0.006 | 0.016 |
| HD45415 | 0.015 | 0.011 | 0.008 |
| HD46116 | 0.009 | -0.004 | 0.004 |
| HD53629 | 0.017 | 0.020 | 0.020 |
| HD54131 | 0.012 | 0.003 | 0.000 |
| HD56160 | 0.010 | 0.033 | 0.025 |
| HD60060 | 0.018 | 0.012 | 0.009 |
| HD60341 | 0.010 | 0.021 | 0.020 |
| HD62412 | 0.013 | 0.013 | 0.010 |
| HD62713 | 0.005 | 0.010 | 0.014 |
| HD68312 | 0.011 | -0.009 | 0.007 |
| HD74622 | 0.013 | 0.023 | 0.015 |
| HD75916 | 0.008 | 0.071 | 0.018 |
| HD176704 | 0.007 | 0.057 | 0.026 |
| HD177873 | 0.002 | 0.004 | 0.035 |

HD191584  0.009  0.032  0.028
HD219784  0.001  0.010  0.016
HD220572  0.003  0.005  0.015
HD204381  0.001   ---   0.012

Except for very few individual stars, the overall agreement of the results is very good. $E(B - V)$ is the reddening adopted in our work based on the maps of Schlegel[24]. $E(B - V)_1$ is based on observed colours and effective temperature, while $E(B - V)_2$ is from fitting SEDs.

**Extended Data Table 4. Contributions to the total statistical errors**

| Name | σE(B-V) | σMC | σtl |
|---|---|---|---|
| ECL-01866 | 0.010 | 0.016 | 0.025 |
| ECL-03160 | 0.012 | 0.015 | 0 |
| ECL-05430 | 0.006 | 0.010 | 0 |
| ECL-06575 | 0.009 | 0.006 | 0 |
| ECL-09114 | 0.016 | 0.014 | 0.018 |
| ECL-09660 | 0.017 | 0.008 | 0 |
| ECL-09678 | 0.013 | 0.013 | 0 |
| ECL-10567 | 0.007 | 0.009 | 0.008 |
| SC9_230659 | 0.015 | 0.019 | 0.01 |
| ECL-12669 | 0.014 | 0.013 | 0 |
| ECL-12875 | 0.024 | 0.011 | 0 |
| ECL-12933 | 0.021 | 0.014 | 0 |
| ECL-13360 | 0.012 | 0.006 | 0 |
| ECL-13529 | 0.012 | 0.011 | 0 |
| ECL-15260 | 0.022 | 0.026 | 0 |
| ECL-18365 | 0.014 | 0.015 | 0 |
| ECL-18836 | 0.011 | 0.016 | 0 |
| ECL-25658 | 0.012 | 0.010 | 0 |

We present (left to right) the name of the system, error on the extinction, error estimated from the Monte Carlo simulations (light curves, radial velocity curves, colour disentangling), and the error from the contribution of the third light.

**Acknowledgements** The research leading to these results has received funding from the European Research Council (ERC) under the European Union's Horizon 2020 research and innovation programme (grant agreement no. 695099). We acknowledge support from the IdP II 2015 0002 64 and DIR/WK/2018/09 grants of the Polish Ministry of Science and Higher Education. We also gratefully acknowledge financial support for this work from the BASAL Centro de Astrofisica y Tecnologias Afines (CATA, AFB- 70002), and from the Millennium Institute for Astrophysics (MAS) of the Iniciativa Milenio del Ministerio de Economía, Fomento y Turismo de Chile, project IC120009. We also acknowledge support from the Polish National Science Center grant MAESTRO DEC-2012/06/A/ST9/00269. We acknowledge the support of the French Agence Nationale de la Recherche (ANR), under grant ANR-15-CE31-0012-01 (project

UnlockCepheids). S.V. gratefully acknowledges the support provided by Fondecyt reg. no. 1170518. This work is based on observations made with ESO telescopes under programmes 092.D-0297, 094.D-0074, 098.D-0263(A,B), 097.D-0400(A), 097.D-0150(A), 097.D-0151(A) and CNTAC programmes CN2016B-38, CN2016A-22, CN2015B-2 and CN2015A-18. This research was supported by the Munich Institute for Astro- and Particle Physics (MIAPP) of the DFG cluster of excellence "Origin and Structure of the Universe".